\documentstyle[12pt]{article}
\begin{document}

\begin{flushright}
INFNNA-IV-99/24
\end{flushright}
\vspace*{1.cm}
\begin{center}
{RARE KAON DECAYS\footnote{\small Work supported in part
by TMR, EC--Contract No. ERBFMRX--CT980169 (EURODA$\Phi$NE)}}\\
\vspace*{1.6cm}
{Giancarlo D'Ambrosio \\
\vspace*{0.4cm}
{\em INFN-Sezione di Napoli, 80126 Napoli Italy}}
\vspace*{1.cm}
\date{30 May 99}
\vspace*{1.6cm}
\begin{abstract}
We review some recent theoretical results on rare kaon decays. Particular
attention is devoted to establish the short distance (direct CP violating)
contribution to $K_{L}\rightarrow \pi ^{0}e\overline{e}$. This is achieved
by a careful study of the long distance part. As byproduct, we discuss
interesting chiral tests.
\end{abstract}
\end{center}

\section{Introduction}

Rare Kaon decays are going to play undoubtedly a crucial role to test the
Standard Model (SM) and its extensions \cite{reviews}. $K_{L}\rightarrow \pi
^{0}\nu \overline{\nu }$ and $K^{\pm }\rightarrow \pi ^{\pm }\nu \overline{
\nu }$ have the advantage not to be affected by long distance uncertainties
and thus they are definetely very appealing \cite{buchalla96,gino98a};
however due to the difficult detection they represent a real experimental
challenge. There are also other interesting channels that can be studied
either as byproduct of the previous ones or also as an independent search,
like $K_{L}\rightarrow \pi ^{0}e\overline{e}$ or $K_{L}\rightarrow \mu 
\overline{\mu }.$ Here the long distance contributions is in general not
negligible and must be carefully studied in order to pin down the short
distance part. The advantage is that these channels are more accessible
experimentally.

\section{$K\rightarrow \pi \nu \overline{\nu }$}

The basic formalism to study systematically low energy physics is the OPE
(Operator Product Expansion), where the physical processes are determined by
an effective hamiltonian, written as a product of local operators $O_{i}$
and (Wilson) coefficients $c_{i}:$ ${\cal H}_{eff}=\sum_{i}$ $c_{i}(\mu )$ $%
O_{i}(\mu );$ indeed the scale dependence must cancel in the product since
physical processes are $\mu -$independent. For $K\rightarrow \pi \nu 
\overline{\nu }$ this writes as \cite{buchalla96}

\begin{equation}
{\cal H}_{eff}=\frac{\displaystyle G_{F}}{\displaystyle \sqrt{2}}\,\frac{%
\displaystyle 2\alpha }{\displaystyle \pi \sin ^{2}\Theta _{W}}\displaystyle %
\sum_{l=e,\mu ,\tau }\left( \lambda _{c}X_{NL}^{l}+\lambda
_{t}X(x_{t})\right) \overline{s}_{L}\gamma ^{\mu }d\,_{L}\overline{\nu }
_{L}\gamma _{\mu }\nu _{L}+H.c.  \label{eq:heffkpnn}
\end{equation}
where $\lambda _{q}=\,V_{qs}^{*}V_{qd}$ and $X(x_{t})$, $X_{NL}^{l}$ the
box+Z-penguin top and charm loop contributions. The latter is affected by
strong radiative corrections. $SU(2)$ isospin symmetry relates hadronic
matrix elements for $K\rightarrow \pi \nu \overline{\nu }$ to $K\rightarrow
\pi l\overline{\nu }$ $.$ QCD corrections have been evaluated at
next-to-leading order and the main uncertainties in (\ref{eq:heffkpnn}) is
due to the strong corrections to $X_{NL}^{l}$, that implies $5\%$ error on
the determination of $\lambda _{t}$ from $K^{^{\pm }}\rightarrow \pi ^{\pm
}\nu \overline{\nu }$.

The structure in $($\ref{eq:heffkpnn}) leads to a pure CP violating
contribution to $K_{L}\rightarrow \pi ^{0}\nu \overline{\nu },$ induced only
from the top loop contribution and thus proportional to $\Im m(\lambda _{t})$
and free of hadronic uncertainties. This leads to the prediction \cite
{buchalla96}

\begin{equation}
B(K_{L}\rightarrow \pi ^{0}\nu \overline{\nu })_{SM}=4.25\times
10^{-10}\left[ \frac{\bar{m}_{t}(m_{t})}{170GeV}\right] ^{2.3}\left[ \frac{
\Im m(\lambda _{t})}{\lambda ^{5}}\right] ^{2}.
\end{equation}

On the other hand $K^{^{\pm }}\rightarrow \pi ^{\pm }\nu \overline{\nu }$
receives CP conserving and CP violating contributions proportional
respectively to $\Re e(\lambda _{c})X_{NL}^{l}+\Re e(\lambda _{t})X(x_{t})$
and $\Im m(\lambda _{t})X(x_{t}).\;$If one takes into account the various
indirect limits on CKM elements one obtains \cite{reviews, buchalla96}

\begin{equation}
B(K_{L}\rightarrow \pi ^{0}\nu \overline{\nu })_{SM}\ =\left( 2.8\pm
1.1\right) \times 10^{-11}
\end{equation}

\begin{equation}
B(K^{\pm }\rightarrow \pi ^{\pm }\nu \overline{\nu })_{SM}=\left( 0.79\pm
0.31\right) \times 10^{-10}
\end{equation}
To be compared respectively with the experimental results \cite{kpi0pi0nunu}
and \cite{kpipnunu} 
\begin{equation}
B(K_{L}\rightarrow \pi ^{0}\nu \overline{\nu })\ \leq 5.9\times 10^{-7}{\
\qquad }B(K^{^{\pm }}\rightarrow \pi ^{\pm }\nu \overline{\nu })=\left(
4.2_{-3.5}^{+9.7}\right) \times 10^{-10}
\end{equation}
Lately it has been pointed out the possibility of new physics to
substantially enhance the SM\ predictions ($\sim $a factor $10$ in the
branching) \cite{gino98b} through effects that could be parametrized by an
effective $\overline{s}dZ$ vertex $Z_{ds};$ the CP violating contribution $%
\Im m(Z_{ds}) $ and consequentely $K_{L}\rightarrow \pi ^{0}\nu \overline{%
\nu }$ is constrained by the value of $\varepsilon ^{\prime }/\varepsilon $,
while $\Re e(Z_{ds})$ and $K^{^{\pm }}\rightarrow \pi ^{\pm }\nu \overline{
\nu }$ are limited by $K_{L}\rightarrow \mu \overline{\mu }.\;$Indeed it has
been shown that it is possible to evaluate in a realiable way the long
distance contributions to $K_{L}\rightarrow \mu \overline{\mu }$ and thus to
constrain the short distance part \cite{klmumu}.

The recent value of $\varepsilon ^{\prime }/\varepsilon $ \cite{KTeVeps99},
though not incompatible with the SM, allows large values for new sources of
CP violating contributions.

\section{$K_{L}\rightarrow \pi ^{0}e^{+}e^{-}$}

\subsection{Direct CP violating contributions}

The electromagnetic interactions allow new structures to this decay compared
to $K\rightarrow \pi \nu \overline{\nu }$ .\ The effective Hamiltonian for $%
s\to d\ell ^{+}\ell ^{-}$ transitions is known at next--to--leading order
and for $\mu <m_{c}$ reads \cite{BL94} 
\begin{equation}
{\cal H}_{eff}^{|\Delta S|=1}=\frac{\displaystyle G_{F}}{\displaystyle \sqrt{
2}}\,V_{us}^{*}V_{ud}\Big[ \,\sum_{i=1}^{6,7V}(z_{i}(\mu )+\tau y_{i}(\mu
))Q_{i}(\mu )~+~\tau y_{7A}(M_{W})Q_{7A}(M_{W})\,\Big]\,+\,\mbox{h.c.},
\label{eq:heffkpee}
\end{equation}
where $\tau =-(V_{ts}^{*}V_{td})/(V_{us}^{*}V_{ud})$ , $Q_{1,..,6}$ are the
usual four-quarks operators and 
\begin{equation}
Q_{7V}\,=\,\overline{s}\gamma ^{\mu }(1-\gamma _{5})d\,\overline{\ell }
\gamma _{\mu }\ell \qquad \mbox{and}\qquad Q_{7A}\,=\,\overline{s}\gamma
^{\mu }(1-\gamma _{5})d\,\overline{\ell }\gamma _{\mu }\gamma _{5}\ell
\end{equation}
are generated by electroweak penguins and box diagrams.

Thus there is i) a direct CP violating contribution analogous to (\ref
{eq:heffkpnn}) (additional single photon exchange contributions are
smaller), ii) indirect $CP$ --violating contribution $K_{L}=K_{2}+\tilde{
\varepsilon}K_{1}\stackrel{\tilde{\varepsilon}K_{1}}{\rightarrow }\pi ^{0}%
\stackrel{*}{\gamma }\rightarrow \pi ^{0}e^{+}e^{-}$ and iii) a $CP-$
conserving contribution: $K_{L}\rightarrow \pi ^{0}\stackrel{*}{\gamma }%
\stackrel{*}{\gamma }\rightarrow \pi ^{0}e^{+}e^{-}.$

The prediction for the direct CP\ violation contribution is \cite{BL94}

\[
B(K_{L}\rightarrow \pi ^{0}e^{+}e^{-})_{CPV-dir}^{SM}\ =0.69\times
10^{-10}\left[ \frac{\bar{m}_{t}(m_{t})}{170GeV}\right] ^{2}\left[ \frac{\Im
m(\lambda _{t})}{\lambda ^{5}}\right] ^{2}; 
\]
using the present constrains on $\Im m(\lambda _{t})$ one obtains \cite
{reviews} 
\[
B(K_{L}\rightarrow \pi ^{0}e^{+}e^{-})_{CPV-dir}^{SM}=\left( 4.6\pm
1.8\right) \times 10^{-12}. 
\]

\subsection{Indirect CP violating contribution, $K_{S}\rightarrow \pi
^{0}e^{+}e^{-}$ and $K^{\pm }\rightarrow \pi ^{\pm }l^{+}l^{-}$}

The interplay between long and short distance contributions to $K\rightarrow
\pi \stackrel{*}{\gamma }$ is manifest in (\ref{eq:heffkpee}), where the
Wilson coefficient of the $Q_{7}-$operator is scale-dependent while the $%
Q_{7}-$matrix element is not.\ Thus the scale-dependence must be cancelled
by the four-quark operators in (\ref{eq:heffkpee}), which have large long
distance (non-perturbative) contributions.

Chiral Symmetry is the appropriate framework to evaluate these
contributions.\ Chiral pertubation theory ($\chi PT)$ \cite{Weinberg1} is an
effective field theory based on the following two assumptions: i) the
pseudoscalar mesons are the Goldstone bosons (G.B.) of the symmetry $%
SU(3)_{L}\otimes SU(3)_{R}$ spontaneously broken to $SU(3)_{V}$ , ii) there
is a {\it (chiral) power counting , }i.e. the theory has a small expansion
parameter: $p^{2}/$ $\Lambda _{\chi SB}^{2}$ and/or $m^{2}/$ $\Lambda _{\chi
SB}^{2},$ where $p$ is the external momenta, $m$ the masses of the G.B.'s
and $\Lambda _{\chi SB}$ is the chiral symmetry breaking scale: $\Lambda
_{\chi SB}\sim 4\pi F_{\pi }\sim 1.2\;GeV$. Being an effective field theory,
loops and counterterms are required by unitarity and have to be evaluated
order by order.

$K\rightarrow \pi \stackrel{*}{\gamma }$ ($K^{\pm }\rightarrow \pi ^{\pm}%
\stackrel{*}{\gamma }$ and $K_{S}\rightarrow \pi ^{0}\stackrel{*}{\gamma }$)
decays start at ${\cal O(}p^{4})$ in $\chi $PT with loops (dominated by the $%
\pi \pi -$cut$)$ and counterterm contributions \cite{EPR1}.\ Higher order
contributions (${\cal O(}p^{6}))$ might be large, but are not completely
under control since new (and with unknown coefficients) counterterm
structures appear \cite{r2}.\ In Ref.~\cite{r3} we have parameterized the $%
K\rightarrow \pi \stackrel{*}{\gamma }(q)$ form factor as 
\begin{equation}
W_{i}(z)\,=\,G_{F}M_{K}^{2}\,(a_{i}\,+\,b_{i}z)\,+\,W_{i}^{\pi \pi
}(z)\;,\qquad \qquad i=\pm ,S  \label{eq:ctkpg}
\end{equation}
with $z=q^{2}/M_{K}^{2}$, and where $W_{i}^{\pi \pi }(z)$ is the loop
contribution, given by the $K\rightarrow \pi \pi \pi $ unitarity cut and
completely known up to ${\cal O(}p^{6})$. All our results in that reference
were given in terms of the unknown parameters $a_{i}$ and $b_{i},$ expected
of ${\cal O(}1)$. At the first non-trivial order, ${\cal O(}p^{4}),$ $%
b_{i}=0,$ while $a_{i}$ receive counterterm contributions not determined
yet. At ${\cal O(}p^{6}),$ $b_{i}\neq 0,$ while $a_{i}$ receive new
counterterm contributions. Due to the generality of (\ref{eq:ctkpg}), we
expect that $W_{i}(z)$ is a good approximation to the complete form factor
of ${\cal O}(p^{6})$ .

Experimentally the $K^{+}\rightarrow \pi ^{+}l^{+}l^{-}$ ($l$=$e,\mu )$
widths and $K^{+}\rightarrow \pi ^{+}e^{+}e^{-}$ slope have been measured,
while in the muon channel the slope has been measured only after our paper 
\cite{kpmumuE865}. We can fix $a_{+}$ and $b_{+}$ from the $K^{+}\rightarrow
\pi ^{+}e^{+}e^{-}$ rate and spectrum respectively. Thus we can predict the
ratio ($R)$ of the widths $\mu /e;$ which however overestimates the
experimental findings \cite{r3,r4} by $2.2$ $\sigma $ 's. Due to the
generality of the form factor (\ref{eq:ctkpg}) we thought that the
experimental situation should improve. Indeed data from E865 \cite
{kpmumuE865} in $K^{+}\rightarrow \pi ^{+}\mu ^{+}\mu ^{-}$ confirms our
prediction: i) it is possible to describe well both leptonic channels with (%
\ref{eq:ctkpg}) and $B(K^{+}\rightarrow \pi ^{+}\mu ^{+}\mu ^{-})$ is now $%
3.2$ $\sigma $ 's larger than the previous measurement \cite{r4} and even
more interestingly ii) the fit with (\ref{eq:ctkpg}), i.e. with the genuine
chiral contributions $W_{i}^{\pi \pi }(z),$ is better ($\chi ^{2}$ /$d.o.f.$ 
$\sim 19.9/9)$ than just a linear slope ( $\chi ^{2}$/$d.o.f.$ $\sim \chi
_{\min }^{2}+9),$ showing the validity of the chiral expansion.

There is no model independent relation among $a_{S}$ and $a_{+}$ and thus a
secure determination of $B(K_{L}\rightarrow \pi
^{0}e^{+}e^{-})_{CP-indirect} $ requires a direct measurement of $%
B(K_{S}\rightarrow \pi ^{0}e^{+}e^{-}),$ possibly to be performed by KLOE at
DA$\Phi $NE \cite{r3}. The dependence from $b_{S}$ is very mild and thus we
predict $B(K_{S}\rightarrow \pi ^{0}e^{+}e^{-})\,\simeq
\,5.2\,a_{S}^{2}\,\times 10^{-9}~$.

If we include the interference term among direct and indirect the $CP$
--violating terms we obtain 
\begin{equation}
B(K_{L}\rightarrow \pi ^{0}e^{+}e^{-})_{CPV}\,=\,\left[
15.3\,a_{S}^{2}\,-\,6.8\frac{\displaystyle \Im m\lambda _{t}}{\displaystyle %
10^{-4}}\,a_{S}\,+\,2.8\left( \frac{\displaystyle \Im m\lambda _{t}}{%
\displaystyle 10^{-4}}\right) ^{2}\right] \times 10^{-12}~.
\label{eq:cpvtot}
\end{equation}
A very interesting scenario emerges for $a_{S}\stackrel{<}{_{\sim }}-0.5$ or 
$a_{S}\stackrel{>}{_{\sim }}1.0$. Since $\Im m\lambda _{t}$ is expected to
be $\sim 10^{-4}$, one would have $B(K_{L}\rightarrow \pi
^{0}e^{+}e^{-})_{CPV}\stackrel{>}{_{\sim }}10^{-11}$ in this case. Moreover,
the $K_{S}\rightarrow \pi ^{0}e^{+}e^{-}$ branching ratio would be large
enough to allow a direct determination of $|a_{S}|$. Thus, from the
interference term in (\ref{eq:cpvtot}) one could perform an independent
measurement of $\Im m\lambda _{t}$, with a precision increasing with the
value of $|a_{S}|$.

\subsection{CP conserving contributions: ``$\gamma \gamma "$ intermediate
state contributions}

\hspace*{0.1cm} The general amplitude for $K_{L}(p)\rightarrow \pi
^{0}\gamma (q_{1})\gamma (q_{2})$ can be written in terms of two independent
Lorentz and gauge invariant amplitudes $A(z,y)$ and $B(z,y):$

\begin{eqnarray}
M^{\mu \nu } &=&\frac{\displaystyle A(z,y)}{\displaystyle m_{K}^{2}}%
\,(q_{2}^{\mu }\,q_{1}^{\nu }\,-\,q_{1}\cdot q_{2}g^{\mu \nu })\;+\;\frac{%
\displaystyle 2\,B(z,y)}{\displaystyle m_{K}^{4}}\,(-p\cdot q_{1}\,p\cdot %
q_{2}\,g^{\mu \nu }\,-\,q_{1}\cdot q_{2}\,p^{\mu }p^{\nu }\,  \nonumber \\
&&\qquad \qquad \qquad \qquad \qquad \qquad \qquad \qquad \qquad \;+\,p\cdot %
q_{1}\,q_{2}^{\mu }p^{\nu }\,+\,p\cdot q_{2}\,p^{\mu }q_{1}^{\nu }\,\,) 
\nonumber \\
&&  \label{kpgg}
\end{eqnarray}
where $y=p\cdot (q_{1}-q_{2})/m_{K}^{2}$ and $z\,=%
\,(q_{1}+q_{2})^{2}/m_{K}^{2}$. Then the double differential rate is given
by 
\begin{equation}
\frac{\displaystyle \partial ^{2}\Gamma }{\displaystyle \partial y\,\partial %
z}=\,\frac{\displaystyle m_{K}}{\displaystyle 2^{9}\pi ^{3}}[%
\,z^{2}\,|\,A\,+\,B\,|^{2}\,+\,\left( y^{2}-\frac{\displaystyle \lambda
(1,r_{\pi }^{2},z)}{\displaystyle 4}\right) ^{2}\,|\,B\,|^{2}\,]~,
\label{eq:doudif}
\end{equation}
where $\lambda (a,b,c)$ is the usual kinematical function and $r_{\pi }=m_{%
\pi }/m_{K}$. Thus in the region of small $z$ (collinear photons) the $B$
amplitude is dominant and can be determined separately from the $A$
amplitude. This feature is crucial in order to disentangle the CP-conserving
contribution $K_{L}\rightarrow \pi ^{0}e^{+}e^{-}.$

We must warn however about the danger of the potentially large background
contribution from $K_{L}\rightarrow e^{+}e^{-}\gamma \gamma $ to $%
K_{L}\rightarrow \pi ^{0}e^{+}e^{-}$ \cite{greenlee}$.$

The two photons in the $A$-type amplitude are in a state of total angular
momentum $J=0$ ($J,$ total diphoton angular momentum), and it turns out that
for this contribution $A(K_{L}\rightarrow \pi ^{0}e^{+}e^{-})_{J=0}\sim
m_{e} $ ($m_{e}$ electron mass) \cite{EPR88}; however the higher angular
momentum state $B$-type amplitude in (\ref{kpgg}), though chirally and
kinematically suppressed for $A(K_{L}\rightarrow \pi ^{0}\gamma \gamma ),$
generate $A(K_{L}\rightarrow \pi ^{0}e^{+}e^{-})_{J\neq 0}$ competitive with
the CP violating contributions \cite{r2}.

The leading finite ${\cal O}(p^{4})$ amplitudes of $K_{L}\rightarrow \pi
^{0}\gamma \gamma $ generates only the $A$--type amplitude in Eq.~(\ref
{eq:doudif}). This underestimates the observed branching ratio, $(1.68\pm
0.07\pm 0.08)\times 10^{-6}$ \cite{KTeVkpgg99} by a large factor but
reproduces the experimental spectrum, predicting no events at small $z$. The
two presumably large ${\cal O}(p^{6})$ contributions have been studied: i)
the ${\cal O}(p^{6})$ unitarity corrections \cite{CD93,CE93,KH94} that
enhance the ${\cal O}(p^{4})$ branching ratio by $40\%$ and generate a $B$%
--type amplitude, ii) the vector meson exchange contributions that are in
general model dependent \cite{SE88,EP90} but can be parameterized $%
K_{L}\rightarrow \pi ^{0}\gamma \gamma $ by an effective vector coupling $%
a_{V}$ \cite{EP90} : 
\begin{eqnarray}
A &=&{\frac{G_{8}M_{K}^{2}\alpha _{em}}{\pi }}a_{V}(3-z+r_{\pi }^{2})~, 
\nonumber \\
B &=&-{\frac{2G_{8}M_{K}^{2}\alpha _{em}}{\pi }}a_{V}~,  \label{eq:abvec}
\end{eqnarray}
where $G_{8}$ is the octet weak coupling, determined from $K\rightarrow \pi
\pi $. Thus the contribution to $K_{L}\rightarrow \pi ^{0}e^{+}e^{-}$ is
determined by the value of $a_{V}.$ The agreement with experimental $%
K_{L}\rightarrow \pi ^{0}\gamma \gamma $ rate and spectrum would demand $%
a_{V}\sim 0.9\;$\cite{CE93} .

It would be desirable to have a theoretical understanding of this value.\
Indeed we have related $a_{V}$ with the $K_{L}\rightarrow \gamma
(q_{1},\epsilon _{1})\gamma ^{*}(q_{2},\epsilon _{2})$ slope \cite{DP97}.
The $K_{L}\rightarrow \gamma (q_{1},\epsilon _{1})\gamma ^{*}(q_{2},\epsilon
_{2})$ amplitude, $A_{\gamma \gamma ^{*}}(q_{2}^{2}),$ can be approximated
by a linear slope $A_{\gamma \gamma ^{*}}(q_{2}^{2})\approx A_{\gamma \gamma
}^{exp}\cdot (1+bx)$, where $A_{\gamma \gamma }^{exp}$ is the experimental
amplitude $A(K_{L}\rightarrow \gamma \gamma )$ and $x=q_{2}^{2}/m_{K}^{2}$.
The slope can be estimated \cite{DP97}, $b_{exp}\,=\,0.81\pm 0.18\,$.
Theoretically the slope $b$ is also generated by vector meson exchange
contribution.

We can evaluate now $a_{V}$ and the $K_{L}\rightarrow \gamma \gamma ^{*}$
slope $b$ in factorization (FM), i.e. writing a {\it current}$\times ${\it %
current }structure 
\begin{equation}
{\cal L}_{FM}\;=\;4\,k_{F}\,G_{8}\,\langle \,\lambda \,J_{\mu }\,J^{\mu
}\,\rangle \;+\;h.c.\;\;\;,  \label{FM}
\end{equation}
where $\lambda \,\equiv \,\frac{1}{2}\,(\lambda _{6}\,-\,i\lambda _{7})$ and
the fudge factor $k_{F}\sim {\cal O}(1)$ has to be determined
phenomenogically. A satisfactory understanding of the model would require $%
k_{F}\sim 0.2-0.3,$ to match the perturbative result.

There are two ways to derive the FM weak lagrangian generated by resonance
exchange (this corresponds to different ways to determine the conserved
current $J_{\mu }$) \cite{DP97,DP98}~:

\begin{itemize}
\item[(${\cal A}$)]  To evaluate the strong action generated by resonance
exchange, and then perform the factorization procedure in Eq.~(\ref{FM}). By
this way, since we apply the FM procedure once the vectors have already been
integrated out the lagrangian is generated at the kaon mass scale.

\item[(${\cal B}$)]  Otherwise, we can first write down the spin--1 strong
and weak chiral lagrangian.\ The general effective weak $VP\gamma $
lagrangian contributing to both ${\cal O}(p^{6})$ $K\rightarrow \pi \gamma
\gamma $ and $K_{L}\rightarrow \gamma \gamma ^{*}$ processes \cite{DP97}
writes as 
\begin{equation}
{\cal L}_{W}(VP\gamma )\;=\;G_{8}\,F_{\pi }^{2}\,\varepsilon _{\mu \nu
\alpha \beta }\,\sum_{i=1}^{5}\,\kappa _{i}\,\langle \,V^{\mu }\,T_{i}^{\nu
\alpha \beta }\rangle \;,  \label{eq:mostg}
\end{equation}
where $T_{i}^{\nu \alpha \beta }$ $i=1,..,5$ are all the possible relevant $%
P\gamma $ structures , the $\kappa _{i}$ are the dimensionless coupling
constants to be determined and $F_{\pi }\simeq 93\,$MeV. The brackets in
Eq.~(\ref{eq:mostg}) stand for a trace in the flavour space. We apply
factorization and thus we derive the weak resonance couplings $\kappa _{i}\,$
. We then integrate out the resonance fields so that the effective
lagrangian is generated at the scale of the resonance.
\end{itemize}

In principle the two effective actions do not coincide and phenomenology may
prefer one pattern \cite{DP98}.\ In the case at hand, ${\cal A}$ and ${\cal B%
}$ give different structures, however they both generate a good
phenomenology with one free\ parameter $k_{F}$ , i.e.

\begin{equation}
a_{V}\,\simeq \,-0.72\;\;\;\;\;,\;\;\;\;\;b\,\simeq \,0.8-0.9~,
\end{equation}
but with different value of $k_{F}:$ ${\cal A}$ $\Rightarrow $ $k_{F}=1$
while ${\cal B}$ $\Rightarrow $ $k_{F}=0.2.$ Interestingly this seems to
suggest that the matching should be performed at the resonance scale.

Very interestingly the new data from KTeV \cite{KTeVkpgg99} confirms sharply
our prediction for $a_{V}:$ $a_{V}=-0.72\pm 0.05\pm 0.06$ and show a clear
evidence of events at low z. This turns in a more stringent determination
for the CP conserving contribution to $K_{L}\rightarrow \pi ^{0}e^{+}e^{-}$: 
$1.<B(K_{L}\rightarrow \pi ^{0}e^{+}e^{-})\cdot 10^{12}<4$ \cite{r2,DP97}.

\section{Conclusions}

We think that the recent experimental results in $K-$decays, for instance $%
\varepsilon ^{\prime }/\varepsilon $ and $K_{L}\rightarrow \pi ^{0}\gamma
\gamma ,$ let us hope to make sensitive tests of the SM and of its possible
extensions.\ From the theoretical side we expect an improvement in matching
long and short distance contributions and this would lead to an accurate
determination of low energy parameters, for instance a reliable relation
between $a_{S}$ and $a_{+}$ in (\ref{eq:ctkpg}).

\section{Acknowledgements}

I am happy to thank my collaborators G.Ecker, G.Isidori and J.\ Portol\'{e}s
for the enjoyable work and the organizers of the  
Conference "La Thuile 99" for the stimulating atmosphere.

\end{document}